\documentclass[10pt]{article}

\usepackage{amsmath}
\usepackage{amssymb}

\usepackage{graphicx}

\usepackage{cite}

\usepackage{color} 

\usepackage[labelfont=bf,labelsep=period,justification=raggedright]{caption}

\makeatletter
\renewcommand{\@biblabel}[1]{\quad#1.}
\makeatother

\date{}

\begin{document}

\begin{flushleft}
{\Large
\textbf{Improved 3D Superresolution Localization Microscopy Using Adaptive Optics}
}
\\

Nicolas Piro$^{1}$, 
Thomas Pengo$^{1}$, 
Nicolas Olivier$^{1}$,
 Suliana Manley$^{1,\ast}$
\\
\bf{1} Laboratory for Experimental Biophysics, IPSB, EPFL 1015 Lausanne, Switzerland	
\\
$\ast$ E-mail: suliana.manley@epfl.ch
\end{flushleft}

\section*{Abstract}

We demonstrate a new versatile method for 3D super-resolution microscopy by using a deformable mirror to shape the point spread function of our microscope in a continuous and controllable way. We apply this for 3D STORM imaging of microtubules.

\section*{Introduction}

Super-resolution microscopy techniques are becoming the method of choice to investigate the nanoscopic organization of protein structures in the cellular environment. Point localization based-techniques, such as Stochastic Optical Reconstruction Microscopy (STORM)\cite{Rust2006,Betzig2006,Hess2006} rely on the temporal separation of fluorophores to localized single molecules individually and reconstruct an image based on the sum of the localized positions. These methods are especially interesting, since they provide a 2D localization precision which is only limited by the number of photons emitted by one molecule.

This localization precision can reach $\sim$1nm \cite{Thompson2002,Yildiz2003} by using bright dyes. Moreover, this can be extended into 3D with several methods that differ in terms of performance and complexity~\cite{Shtengel2009,Juette2008,Huang2008,Pavani2009,Grover2011}.

One method is point-spread function (PSF) engineering, where the PSF of the microscope is engineered so that its shape becomes more Z-dependent, and thus encodes the Z position of the molecules. In the context of super-resolution imaging, the first and most commonly used implementation of this scheme made the PSF more or less elliptical as a function of depth by adding a cylindrical lens in the optical path \cite{Huang2008}. The advantage of this method is its simplicity, both optically since only one optical element is added, and computationally, since Gaussian fitting is still  performed, albeit adding an ellipticity parameter to reveal the Z position. However, the cost of this simplicity is the limited performance: since the PSF is not shaped perfectly, information is lost when the Gaussian fitting is performed. Moreover, the tunability is limited, since adding a more strongly focusing cylindrical lens results in more uncontrolled aberrations and does not increase the axial localization. Finally, imaging at large depths is limited by sample induced aberrations, which changes the shape of the PSF in an uncontrolled way. 

One way to limit the influence of the aberrations is to correct the wavefront using adaptive optics. Indeed, adaptive optics is now an established field of microscopy \cite{Booth2007}. It relies on the use of an adaptive element (typically a deformable mirror or a spatial light modulator) to manipulate the phase of either the exciting beam in nonlinear microscopy, or the detected signal in linear imaging. The main use of adaptive optics is to correct the optical aberrations that are introduced by the inhomogeneities of the optical index in the sample. Additionally, it can be used to introduce controlled aberrations to shape the PSF of the microscope. Adaptive optics can therefore be used in 3D STORM to both enhance the localization precision and allow the imaging of thick samples. Since most super-resolution imaging is performed on cells, there are few aberrations introduced by the sample (as long as a correct objective is used), so we focus in this paper on the shaping aspect. We show that using a deformable mirror for PSF shaping in 3D super-resolution  provides high photon efficiency, and high tunability. A previous work described a proof of principle experiment of localization microscopy using a similar approach~\cite{Izeddin2012} but we present 3D STORM imaging of biological samples using a deformable mirror on a simple setup. We also introduce a new algorithm for rendering 3D data in a more physically correct manner.

\section*{Results}

\subsection*{Setup}

We built an inverted fluorescence microscope in a modified 4-f configuration (Figure 1) using standard low-cost optomechanical components. Excitation light from a 635nm laser is focused through a dichroic mirror onto the back focal plane of an oil immersion objective lens. Fluorescence emission from the sample is collected by the objective and transmitted through the dichroic mirror, magnified by a telescope system (x1.67) and reflected by a deformable mirror placed at the conjugate image plane of the back focal plane of the objective. The telescope is designed to match the pupil diameter of the objective (6 mm) to the effective diameter of the deformable mirror (11 mm). The light reflected from the mirror is then directed through a de-magnifying telescope (x0.75) and imaged by the tube lens onto an EMCCD camera, resulting in a pixel size of 100nm. In this configuration, the deformable mirror is located at the Fourier plane of the objective, such that shaping the mirror directly allows engineering of the wavefront of the fluorescence light.

We used a membrane-based deformable mirror equipped with 48 electrostatic actuators, which came pre-calibrated using a wavefront sensor. The influence matrix provided represents the deformation acquired by the mirror upon applying a fixed voltage to each actuator individually. The pseudo-inverse of this influence matrix is a map to convert the desired mirror shape into the set of actuator voltages that must be applied. The mirror is capable of generating any low order Zernike mode with high amplitude and good accuracy. 

\subsection*{Calibration}

2D super-resolution localization microscopy works by making individual fluorophores in the sample stochastically switch between fluorescent and non-fluorescent states such that in any given frame a sparse subset of them is imaged, and their PSFs do not overlap, allowing each fluorophore to be  localized individually. In our setup, 3D super-resolution localization microscopy is achieved by shaping the deformable mirror with the primary astigmatism $Z_2^2$ mode, shown in the inset of Figure 2b. This controlled astigmatism induces an ellipticity in the PSFs of individual fluorophores that depends on their depths in the sample. In this way, the ellipticity of the imaged PSF contains the information of the depths of each fluorophore. Hence, by fitting a model elliptical PSF to each single emitter it is possible to extract their X, Y and Z localizations with resolutions better than the diffraction limit.


To calibrate the system, we recorded the 3D PSF by taking a Z-stack of fluorescent microspheres. Plotting X and Y profiles as a function of the depth Z of the sample (Figure 2b), clearly illustrates the degree of astigmatism present in the optical system. We then fitted elliptical 2D Gaussian functions to each PSF to obtain the width and height of the PSF for each Z position. By then fitting the difference in width and height of the PSF as a function of Z to a quadratic polynomial function, we obtained a calibration curve that can be used to localize the depth of single fluorophores in subsequent super-resolution imaging runs of biological samples. 

We measured three calibration curves for three different amplitudes of the astigmatic mode applied to the mirror (Figure 2b), showing that we can control the amount of astigmatism in our system without moving or replacing any optical element, thus making this setup advantageous as compared to standard astigmatism-based 3D super-resolution microscopes. 

\subsection*{3D Imaging}

We then demonstrated the enhancement in resolution provided by our microscope by imaging the microtubules in a fixed cell. Microtubules are a well characterized structure of $\sim$25nm diameter, that are often used to test the resolution in SR imaging. Since we used antibody labeling against $\alpha$-tubulin (primary + secondary), the expected diameter is in the order of 50nm\cite{Huang2008}.

A 3D image of the sample was obtained by plotting the detected molecule density, encoded in the pixel intensity, as a function of X and Y position, while the Z position was encoded with color (Figure 3a). We adapted the standard rendering to better represent overlapping features within the images, by introducing an attenuation model to account for the expected non-transparency of the structures (see Materials and Methods). Vertical crossing of overlapping microtubules is clearly resolved by the image, as shown in the zoomed regions (Figures 3b-c). To assess the resolution of our image, we plotted histograms of detected molecule counts as a function of X, Y and Z positions for three different sections of microtubules depicted in Figure 3a with fixed orientations along these three axes. We then fitted these histograms with molecular density distributions obtained by the convolution of a top hat function with the expected width of the immunostained microtubules, and a Gaussian function modeling the point-spread-function of the optical system (Figures 3d-f). The fitted width of the Gaussian function in this model is a good estimate of the imaging resolution of our setup. We obtain FWHM sizes of $\sim$40 nm in X and Y, and 90 nm in Z. In contrast to other methods used in the literature to assess image resolution, our method accounts for all different sources of uncertainty, including localization precision, mechanical drift, vibrations and labeling accuracy.

\section*{Discussion}

There are two instrument-related aspects fundamentally limiting the resolution in PSF-shaping based 3D STORM: the number of photons detected per molecule, and the amount of information encoded in the shape. Using a deformable mirror allows us to optimize both points. Indeed, as opposed to spatial light modulators, metallic membrane-based deformable mirrors yield low losses. The mirror used in this work is silver coated, and therefore displays a reflectivity at the wavelength range of interest ($\sim$700nm) of above 95$\%$.

Moreover, using a deformable mirror provides two advantages regarding information encoding. Firstly, the strength of astigmatism applied is tunable, as shown in Figure 2, which means that the PSF can be optimized for each application, depending on its requirements (depth of field, lateral resolution, axial resolution). The slopes of the calibration curves in Figure 2 determine the Z localization precision that can be obtained in a 3D super-resolution image. An increase in Z localization precision is clearly accompanied by a cost in X and Y localization precision due to a higher spread in the X and Y widths of the PSF. Thus, depending on the structure of the sample to be imaged, it is useful to vary the amount of astigmatism applied, in order to obtain the best compromise between X, Y and Z localization precisions. It is also possible to devise strategies to increase image resolution in all three dimensions. For instance, one could take several data sets using different amounts of astigmatism and use an appropriate algorithm to combine them and optimize localization precision in all dimensions.

Secondly, the use of a deformable mirror allows more precise shaping of the PSF. Indeed, using a cylindrical lens does not induce perfect astigmatism, which can easily be seen in the star-shape of the PSF close to the focal plane. Using a deformable mirror, however, perfect astigmatism can be induced, producing more gaussian-like PSFs. When using standard gaussian fitting routines to localize fluorophores, this amounts to an effective increase of the localization precision in all three dimensions, and a reduction of spurious localizations leading to background noise in the final image. 

Apart from these two advantages, specific to localization-based 3D super-resolution microscopy, this setup provides a standard adaptive-optics playground to apply aberration compensation techniques to enhance image resolution in 3D super-resolution microscopy. Indeed, aside from the desired controlled astigmatism that we apply to obtain 3D information in the image, other unwanted aberrations are present in the optical system which lead to a degradation in the quality of the PSFs of individual fluorophores, and hence to image resolution. Measuring these aberrations in some way and shaping the mirror to compensate for them is a clear future application of this system. 

An image-based aberration quantification strategy would naturally fit together with the super-resolution image acquisition sequence. For instance, one could use an appropriate search algorithm based on the optimization of an image quality parameter, such as image sharpness, to find the optimal mirror shape. Most aberrations in microscopy are sample-dependent, due to variations in the optical index of the mounting medium. Thus, this setup provides a versatile way of correcting for these sample-induced aberrations, removing the need for the experimenter to prepare samples with very specific optical properties.

\section*{Materials and Methods}

\subsection*{Microscope Setup}

The system is an inverted fluorescence microscope built using standard optomechanical components (Thorlabs) (Schematics: Figure 1). Lasers at different wavelengths (Coherent Sapphire 532 nm) and (Coherent Cube 635 nm), which can deliver maximum powers of 50 mW and 35 mW respectively, are combined and focused by a 70 mm lens (not shown) onto the back focal plane of an oil immersion objective lens (Olympus UPlanSApo 100x NA=1.4), after a reflection on a two-color dichroic mirror (Z532/633 RPC-XT, Chorma). The sample is placed on a 3-axis piezo-driven translation stage (Thorlabs NanoMax MAX311D) equipped with strain sensors on each axis. While X and Y axes are positioned in open loop mode, the Z axis is controlled in a closed feedback loop, providing high resolution and repeatability (specified to 5 nm). Fluorescence from the sample is collected by the objective, transmitted through the dichroic mirror, filtered by an emission filter (ET700/75, Chroma), magnified by a telescope system (lenses $f_1=150$ mm and $f_2= 250$ mm) and reflected on a deformable mirror (Adaptica Saturn), placed at the conjugate image plane of the back focal plane of the objective. The telescope is designed to match the pupil diameter of the objective (6 mm) to the effective diameter of the deformable mirror (11 mm). The light reflected from the mirror is then directed through a demagnifying telescope ($f_3=200$ mm and $f_4= 150$ mm) and imaged by the tube lens ($f_t=180$ mm) onto an electron-multiplying charged coupled device (EMCCD, Andor iXon 897). In this configuration, the deformable mirror is located at the Fourier plane of the object, such that shaping the mirror directly allows engineering the wavefront of the fluorescence light. 

\subsection*{Calibration}

To calibrate the system, we imaged TetraSpeck (Invitrogen) fluorescent microspheres fixed on the surface of a size1 25mm coverslip (Menzell) using Mowiol (Sigma) as a function of depth to obtain the 3D PSF of the system. We then fitted elliptical 2D Gaussian functions to each PSF to obtain the width and height (FWHM) of the PSF for each Z position. By then fitting the difference in width and height of the PSF as a function of Z to a quadratic polynomial function (Matlab), we obtained a calibration curve that can be used to localize the depth of single fluorophores, in subsequent super-resolution imaging of biological samples.

\subsection*{Sample Preparation}

Cos7 cells were cultured in DMEM (Invitrogen) supplemented with 10\% FBS (fetal bovine serum, Gibco) in a humidified 5\% CO2 incubator at $37^{\circ}$C. 24h after plating on 25 mm size 1 coverslip (Menzell), cells were pre-extracted in 0.5\% Triton X-100 in BRB80 (80 mM K-PIPES, pH 6.8, 1 mM MgCl$_2$; 1 mM EGTA) supplemented with 4 mM EGTA ($37^{\circ}$C), washed in PBS ($37^{\circ}$C), fixed for 15 min in 0.5\% glutaraldehyde (Applichem) diluted in warm PBS, followed by quenching for 10min in 0.1\% sodium borohydride (NaBH4, Sigma, 452904) in water and further PBS washes. Samples were incubated for 1 h at room temperature with alpha-tubulin antibodies (1:1000 mouse anti-alpha-tubulin (Sigma, T5168) in 1\% BSA - 0.2\% Triton X) followed by washes, and secondary antibody (1:1000 anti-mouse Alexa Fluor 647 F(ab')2 (Life Technologies, A21246) in 1\% BSA - 0.2\% Triton). Immunostained coverslips were either directly imaged or kept in PBS (with 1x penicillin-streptomycin, from 100x stock Gibco, 15140) at $4^{\circ}$C until imaging.

\subsection*{Data Analysis}

Raw data is first analyzed using Peakselector (Courtesy of H. Hess). Briefly, individual molecules are first identified, and then fitted using a 2D elliptic Gaussian function. All molecules with less than 1000 detected photons are discarded. Z position is then extracted from the ellipticity using the calibration curve described previously..

\subsection*{Data Rendering}

Typically, three-dimensional point sets are rendered by encoding the density with intensity and depth with color. However, when two layers overlap, the color is averaged across the different depths at that particular point, so that the crossing point between overlapping structures is rendered with a color corresponding to an intermediate position. To overcome this, we adapt the standard rendering by introducing an attenuation model, which emulates the fact that overlapping non-transparent objects occlude each other, which is an appropriate representation if we assume that the structures we render are non-transparent. The intensity at each layer is the sum of an emission term, proportional to the density of points at a particular layer, and a cumulative term, which is an attenuated version of the intensity of the previous layer, attenuated proportionally to the density in the current layer. If $\rho_{ijk}$ is the estimated density of points at pixel $(i,j,k)$, then the RGB vector (${\bf I}_{ij}^k $)  at XY position $(i,j)$ and $Z$ position $k$ is calculated as:

\begin{eqnarray}
{\bf I}_{ij}^k &=& {\bf I}_{ij}^{k-1}(1-\alpha\rho_{ijk})+\beta{\bf H}^k\rho_{ijk}, k=1\dots z  
\end{eqnarray}

where (${\bf H}^k $)  is the color vector corresponding to level $k$, and $\alpha$ and $\beta$ two constants balancing the attenuation of layer $k-1$ and the contribution of layer $k$, respectively. Two special cases are: $\alpha=0$ and $\beta>0$, which gives the classical non-attenuated color-coded depth rendering, and $\alpha=-1$ and $\beta=0$ gives a black-and-white sum projection of the three-dimensional density. In our experiments we have used $\alpha=0.5$ and $\beta=1$, which give a good balance between attenuation and color. Finally, the number of axial layers z is chosen according to the axial localization precision: if the localization precision is 100nm and the displayed depth range is $1\mu m$, $z = 10$ layers are sufficient to give a sufficient level of color separation between layers. Using too few bins would reduce the depth resolution, while using too many bins would induce color noise, as the error of the $z$ estimation may assign two different colors to two different localization densities corresponding to the same depth. 

\section*{Acknowledgments}

We thank H. Hess for Peakselector, Debora Keller for Sample prep, Alexandre Mougel for preliminary experiments

\bibliography{AO}

\section*{Figure Legends}

\begin{figure}[!ht]
\begin{center}
\includegraphics[width=4.86in]{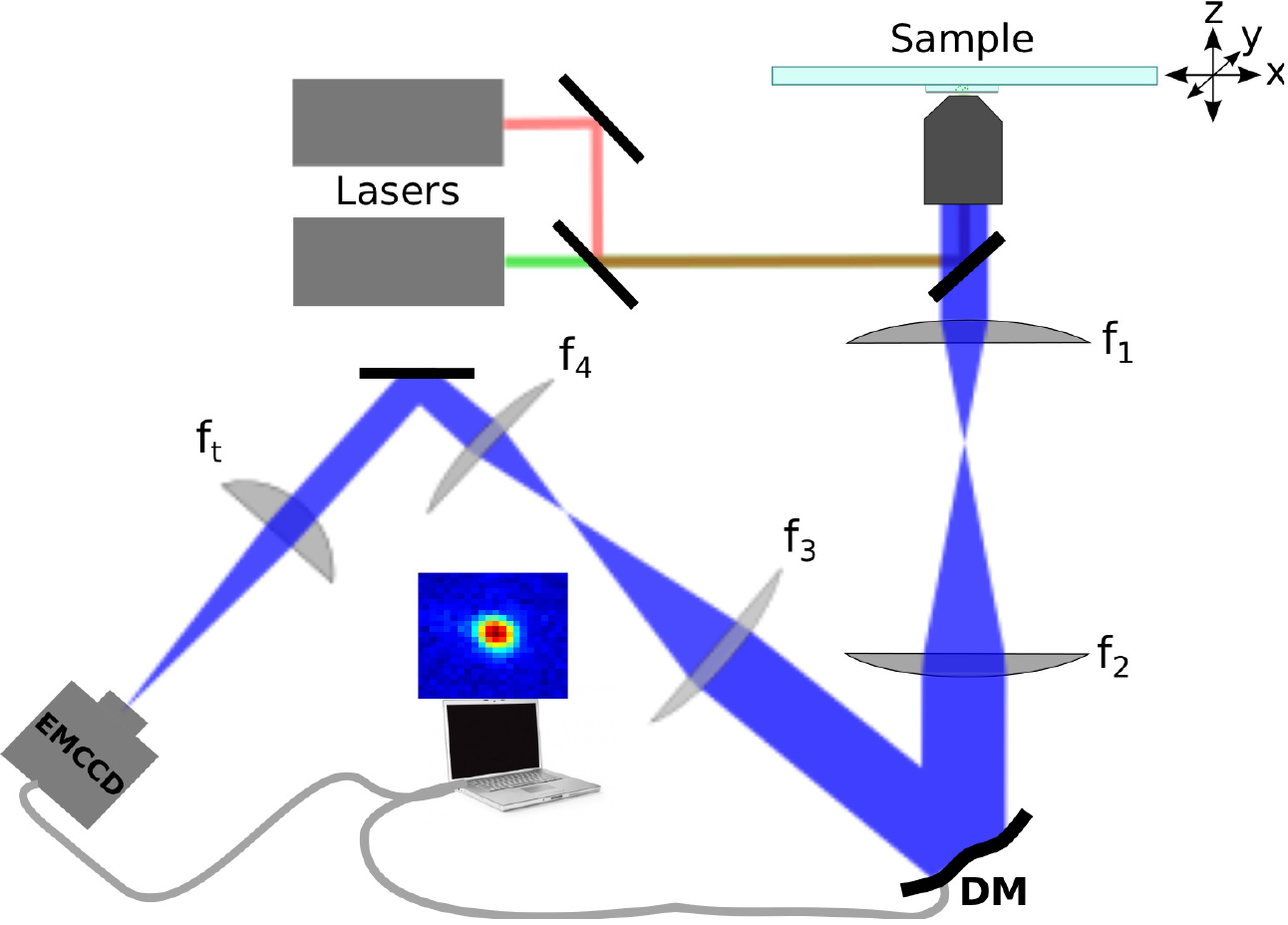}
\end{center}
\caption{
{\bf Optical Setup.} The microscope is built in a 4-f configuration. Lasers at 532 nm and 635 nm  illuminate the sample through the reflection port of a dichroic mirror (the 635 nm laser is used in this work). Fluorescence light from the sample is collected by an oil immersed 100x objective and is transmitted through the dichroic mirror. The back focal plane of the objective lens is magnified by a 1.67x telescope and imaged onto the deformable mirror. A second 0.75x demagnification telescope and the tube lens form an image on an EMCCD camera, such that a single PSF has a diameter (FWHM) of 100 nm.
}
\label{Figure_1}
\end{figure}

\begin{figure}[!ht]
\begin{center}
\includegraphics[width=4.86in]{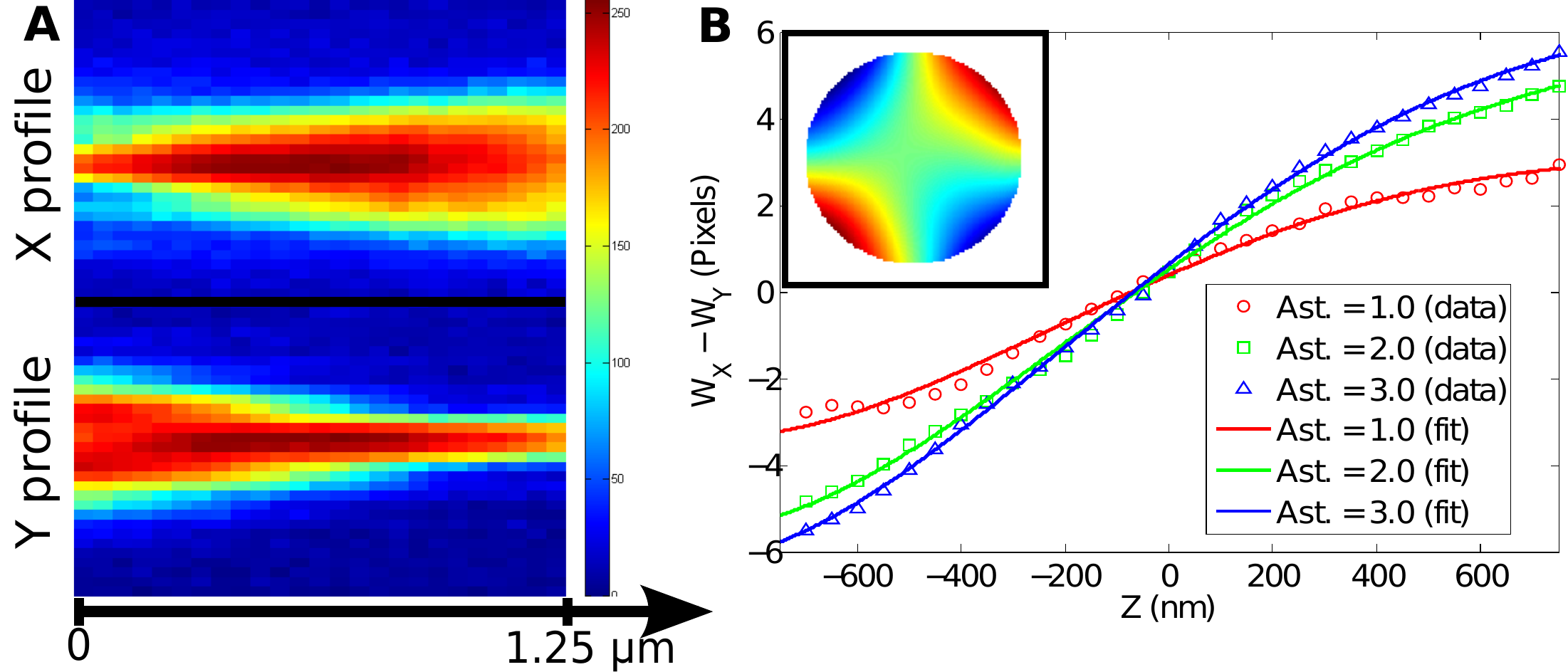}
\end{center}
\caption{
{\bf Controlling the ellipticity of the PSF with astigmatism.}  (a) X and Y vs Z position of the measured PSF, for an applied astigmatism of 2.0. (b) Calibration curve measured with fluorescent beads. The  difference between the horizontal and vertical widths of the PSF vs Z position are plotted together with a fitted forth-order polynomial function, for three different amplitudes of the applied astigmatism. In the inset, the shape applied to the mirror is shown (red, green, blue correspond to positive, null and negative displacements).}
\label{Figure_2}
\end{figure}

\begin{figure}[!ht]
\begin{center}
\includegraphics[width=6.83in]{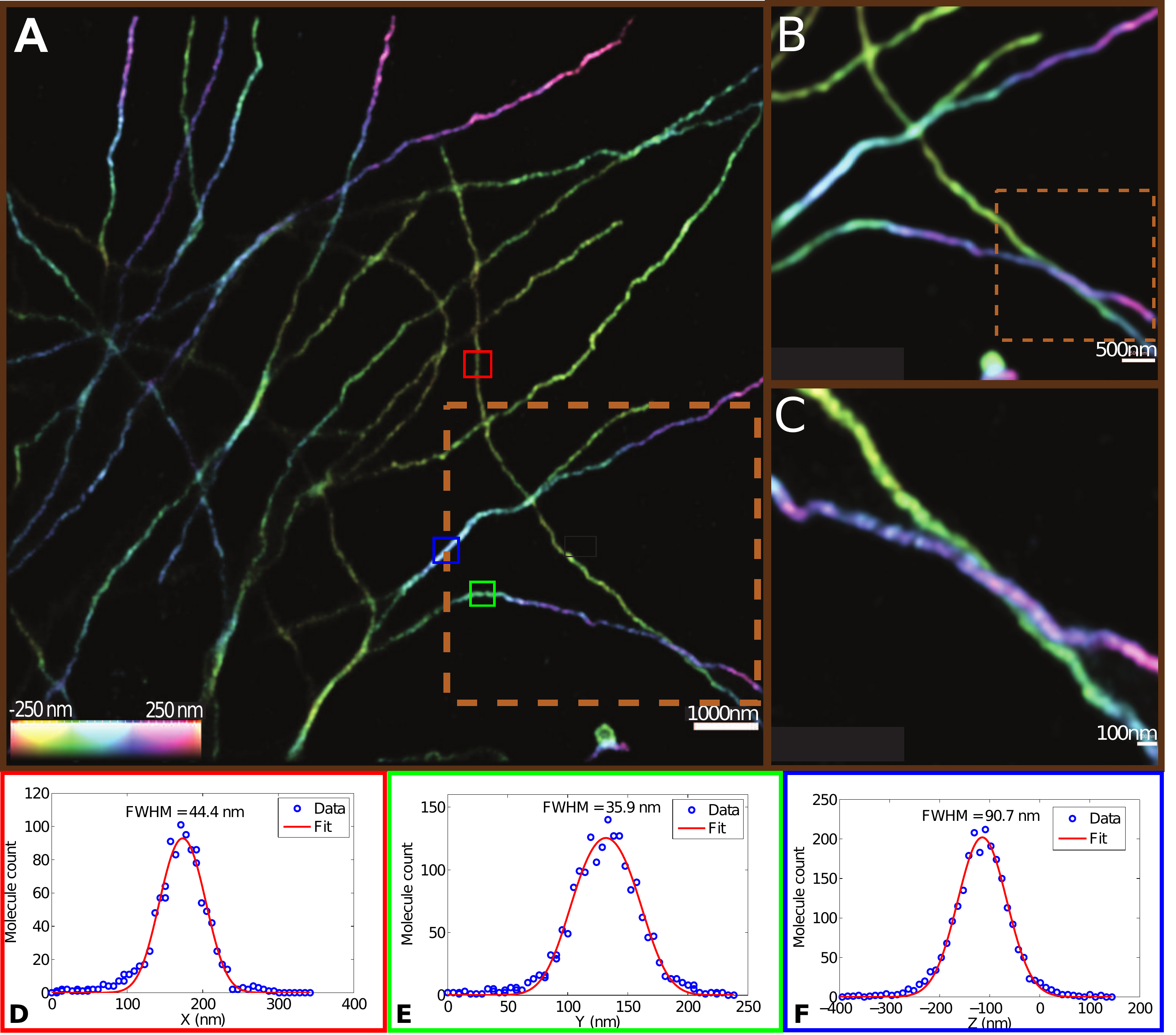}
\end{center}
\caption{
{\bf 3D-STORM imaging of microtubules.}  (a) Full field of view image of the intercrossing of microtubules in a cell. The intensity of each pixel is proportional to the detected density of molecules, while their color is coded to the detected Z position. A value of astigmatism of 2 (conf. Figure 2) was used to obtain this image. The orange rectangular area is zoomed in in (b), and further in (c) showing the resolved twisting of each microtubule and the intercrossing of two of them.  (d)-(f) are histograms of molecule localization in three different regions of the image (red, green and blue, respectively) corresponding to sections where microtubules have constant X, Y and Z orientations. A fit to a model function, formed by convoluting a top hat function with the known width of a microtubule (50 nm) with a gaussian PSF function with its widths as a free parameter, allows us to assess the X, Y and Z imaging resolutions of our microscope.
}
\label{Figure_3}
\end{figure}

\end{document}